\documentstyle[11pt,aaspp]{article}

\nonstopmode	

\def\etal{{\it et al.}}
\singlespace

\begin{document}

\title{Filament Statistics: A Quantitative Comparison of Cold + Hot and
Cold Dark Matter Cosmologies with CfA1 Data}

\author{Romeel Dav\'e$^1$, Doug Hellinger$^2$, \\
        Richard Nolthenius$^1$, Joel Primack$^2$,
             and Anatoly Klypin$^3$
\begin{flushleft}
{\it
$^1$ UCO/Lick Observatory,
     University of California, Santa Cruz, CA 95064   \\
$^2$ Santa Cruz Institute for Particle Physics,
     University of California, Santa Cruz, CA 95064   \\
$^3$ Astronomy Department, New Mexico State University,
     Las Cruces, NM 88001 \\
}
\end{flushleft}
}

\abstract
A new class of geometric statistics for analyzing galaxy catalogs is presented.
{\it Filament statistics} quantify filamentarity and planarity in large
scale structure in a manner consistent with catalog visualizations.
These statistics are based on
sequences of spatial links which follow local high-density
structures.  From these link sequences we compute
the discrete curvature, planarity, and torsion. Filament statistics are applied
to CDM and CHDM ($\Omega_\nu = 0.3$) simulations of Klypin \etal (1995),
the CfA1-like sky catalogs of Nolthenius, Klypin and
Primack (1994, 1995), and the CfA1 catalog.
For 100 Mpc periodic simulation boxes ($H_0 = 50$ km s$^{-1}$ Mpc$^{-1}$),
we find robust discrimination of over 4$\sigma$ (where $\sigma$ represents
resampling errors) between CHDM and CDM.
The {\it reduced filament statistics} show that
CfA1 data is intermediate between CHDM and CDM, but more consistent with
the CHDM models.
Filament statistics provide robust and
discriminatory shape statistics with which to test cosmological simulations
of various models against present and future redshift survey data.


\newpage

\section{Introduction}

We introduce {\it filament statistics}, a new class of geometric
statistics designed to quantify filamentarity and planarity in
large-scale structure.  We compare
cosmological simulations of pure Cold Dark
Matter (CDM) models versus Cold plus Hot Dark Matter (CHDM) models
in real space, as well as
simulated CfA1-like redshift surveys generated from these
simulations versus the CfA1 data.
Visual comparison of the models (Brodbeck \etal 1995; hereafter BHNPK)
shows that the CDM galaxy distribution contains larger clusters and
less well-defined filamentary and sheet-like structures than CHDM.
The filament statistics presented in this paper
confirm as well as quantify these results, showing
statistically significant and robust discrimination between the models.
Filament statistics represent a new and independent class of statistics whose
discriminatory power will likely improve further
as larger and more complete redshift surveys
become available.  We present these initial results to demonstrate the
viability of the methods.

Filament statistics are related to the alignment statistic, originally proposed
by Dekel (1984):  For each galaxy, consider
two concentric shells; find the moment of inertia ellipsoid axes defined by
galaxies
within each shell; and calculate the angle difference between the
inertia tensor axes.  Presumably, where the
angle difference in the major axis is small, there is a filamentary
structure present, and where the angle difference in the minor axis
is small, there is a sheet-like structure present.  By randomly sampling
the galaxy distribution at different shell radii, one can then
gain a measure of the filamentarity
and planarity in large-scale structure at various scale lengths.
We found that the alignment statistic barely discriminated between CDM and CHDM
in real 3D space, and failed to discriminate the models in redshift space.

Our improvement, and the crux of this paper, is to apply generalizations
of this statistic in a {\it geometric network} construction (see Hellinger
\etal 1995).
Filament statistics represent a basic implementation
of this more general class of statistics.  Geometric networks generalize
topological networks, where the galaxies are vertices of graphs, by
considering the configurations of ordered point sets with related statistical
geometric properties.
The visualizations of BHNPK show marked differences in
the number, size, and continuity of filamentary structures in CHDM and CDM.
This inspired us to consider the geometric networks generated by mapping the
point set of galaxies into another point set
by a prescription that sensitively favors contiguous high density regions.
The prescription we employ here is designed to respect the discrete geometric
and topological properties of the data; that is, we work with the original
point
sets of data, either in real space or redshift space.
We define three simple statistics to compute on these geometric networks
which measure filamentarity and planarity.

The next section describes the algorithm and parameter choices
for constructing the geometric network used in filament statistics,
and defines the individual statistics.
The third section outlines the data on which these statistics were calculated.
The fourth section presents results and interpretations.  The
last section describes future work and further applications of
filament statistics.

\section{Implementation of Filament Statistics}

\subsection{The Creation of Link Sequences}

The basis of the geometrical network construction used in filament statistics
is the creation of {\it link sequences} which follow
local high density regions, as determined by the principal axis
of the local moment of inertia tensor.
A link sequence is an ordered set of points which can be visualized as
joined by ``links", created by the procedure outlined in the flowchart
in Figure~\ref{fig: dg_flowchart_filament}.
A link sequence is started from each galaxy in a catalog of galaxies
(or if there are too many, a random subset of such galaxies).
The moment of inertia tensor is computed using the masses and positions
of galaxies within a range $R$ of the given point; for redshift survey data,
we weight by luminosity instead of mass.
The eigenvectors and eigenvalues of the inertia tensor are found, and from
these
the principal axis is determined.
The new point in the sequence is created at a distance $L$ (the ``link length")
away in the direction of the principal axis, and a link is created
which joins the old point to the new point.
Note that only the first point in a link sequence is a galaxy; the others
are simply points within the catalog volume.
A new inertia tensor is computed around
this new point, and the procedure is repeated until termination.
Sequence termination occurs when there
are too few nearby galaxies to reasonably identify an axis.
By this prescription, each galaxy generates a sequence of links.
If a sequence has enough links,
then statistics are computed on this link sequence, otherwise
the sequence is discarded.
The construction of a link sequence is completely defined by choosing the
link length $L$,
the maximum radius of galaxies to be included in computation of the
moment tensor $R$, and the criteria for termination of a sequence.

\subsection{Constructing a Dimensionless Statistic}

We would like to construct dimensionless parameters which describe the
shapes of structures.  For that we need to express all scales in
units of some typical length scale of the catalog.
A natural choice is the mean intergalactic spacing
$\bar d \equiv (V/N)^{1/3}$, where $V$ is catalog volume and $N$
is number of galaxies in the catalog, since it provides
a length scale without any information about shape or clustering; it is
also the simplest choice.

We will consider applications in real space as well as magnitude-limited
and volume-limited redshift space.
Redshift distortion will produce measurable effects on link sequences,
and attempts will be made to understand and quantify these effects.
Whereas in real space the mapping of a galaxy into a link sequence
is completely well-defined, in redshift space this is no longer true ---
redshift distortion for a given structure depends on the
vantage point chosen to observe the structure.
Comparison
of the properties of the distributions of link sequences in real and redshift
space may provide insights useful for constructing corrections for redshift
distortions, which in turn might provide methods for correcting other
statistics
as well.  We defer these corrections to subsequent research and for now
consider
the simplest statistic, which we show is not adversely affected by
redshift distortion.

A complication arises in computing $\bar d$ in magnitude-limited catalogs,
since the sample incompleteness increases with distance from the Milky Way
origin, making $\bar d$ a function of radius from origin.
A local computation of $\bar d$ around a given sequence point
({\it i.e.,} using $\bar d = (V/N)^{1/3}$ for a local volume
around the given point) will degrade the statistics,
since structure identification will be biased
towards underdense regions where $\bar d$ is large, which is
exactly opposite of what is desired.  Instead, $\bar d$ should be
corrected only for the selection function,
which depends only on the distance
$r$ from the origin.  Since $\phi (L) dL$ is the number density of galaxies
between luminosity $L$ and $L+dL$, we can obtain $\bar d (r)$
for galaxies visible above the magnitude limit as follows:

\begin{equation}
\bar d (r) = \left\lbrack {\int^\infty_{L_{\em lim}(r)} \phi (L) dL}
\right\rbrack^{-1/3}
\label{eq: dgmis}
\end{equation}

\noindent
where $L_{\em lim}(r)$ is the luminosity
of a galaxy with apparent magnitude 14.5 (the CfA1 magnitude limit) at
a distance $r$.  $\phi (L)$ is assumed to have Schecter form

$$\phi (L) dL = \phi^\ast \biggl({L\over L^\ast}\biggr)^\alpha \exp(-L/ L^\ast)
{dL\over L^\ast}.$$

\noindent
The Schecter parameters $\phi^\ast$, $L^\ast$ and $\alpha$
were best-fit to each real and simulated redshift catalog
individually; this procedure is described in Nolthenius \etal
(1994, 1995; NKP94 and NKP95, respectively).
Note that the true distance
$r$ is unknown, and is instead estimated assuming no peculiar
velocities, {\it i.e.} $r \equiv v/H_0$ for a galaxy with radial velocity $v$.
A few blueshifted galaxies
(mostly in Virgo) do end up on the opposite side of the origin, but the
statistics turn out to be insensitive to where these few galaxies are placed.
${\bar d (r)}$ is computed and used as the local
mean intergalactic spacing at each sequence point in the
analysis of magnitude-limited catalogs.

\subsection{Link Parameters}

The first parameter choice we tried was the simplest, with
$L =R = \bar d$.  The virtue of this definition
is that we have a parameter-free statistic, in the sense that the
parameters are all determined from intrinsic properties of the data set.
Unfortunately, statistics derived from constructions with these
``natural'' parameters did not discriminate between models
for reasons that will be clarified below.

For link length $L = \bar d$, but range $R$ left as a free parameter, we obtain
discriminatory statistics; this choice of $L$ appears to work as well as
any other.  However, for $R = \bar d$, and $L$ a free parameter,
we again find little discrimination between models, or
even from a Poisson catalog.  $R = \bar d$ turns out to be
too small to identify a local structure, and is dominated by shot noise.
A larger $R$ will yield more points per sphere, thereby lowering shot-noise
scatter.
Since the $R$ parameter controls the scales of structure being
measured by the statistics, it is interesting and instructive to look at
statistics as
a function of $R$, and the results will be presented that way.

An ``optimal" $R$ for a given catalog and statistic may be identified by
maximizing the discrimination of the given catalog from the Poisson
catalog.  In section 4.5 we will show that this optimization yields
consistent and well-defined $R_{opt}$ values.  We shall call the
statistics at $R = R_{opt}$ the {\it reduced filament statistics}.

\subsection{Termination Criteria}

There are three parameters which set the termination criteria for a
link sequence.  $N_{P,min}$ is the minimum number of galaxies required within a
sphere of radius $R$ for a sequence to continue; $N_{P,min}$ was set to
5 so that the determination of the principal axis would be statistically
meaningful, and so that a sequence would terminate if it was in a
sparse region in the catalog.
$N_{L,max}$ sets the
maximum number of links for a periodic catalog,  and
is set so that the total length of a sequence cannot exceed the length
of the simulation box.  In a redshift survey, the sequence terminates if
it exceeds the catalog boundary.
$N_{L,min}$ sets the minimum number of links for
a sequence to be statistically meaningful.  This was set to 4 links
(the minimum value for
computation of all statistics), but can be increased to explore
more extended structures.  However, since each link is typically
fairly large ($\approx$3 Mpc in the simulations considered, and
$\approx$15 Mpc in the sparser CfA1 catalog), 4 links is already
exploring a reasonably extended scale.

All the termination parameters were varied over fairly wide ranges.
$N_{P,min}$ was varied from 4 to 10 with little change in discrimination
or robustness; any higher, and the shot noise generated from fewer
sequences became significant.  The statistics are
independent of $N_{L,max}$ as long it is above about 10, below which
shot noise from the small number of links becomes significant; it is
about 34 in the periodic simulation boxes.  Variations in $N_{L,min}$
had some effect on the results for real and simulated redshift catalogs,
since for very small values (2 or 3)
shot noise increases from low link sampling, while for a high value (above
10), the number of acceptable sequences decreases so that catalog sampling
shot noise becomes large.
Discrimination was also insensitive to the choice of either a Gaussian,
exponential,
or top hat window function;  we used a top hat for computational efficiency.

Note that the principal axis of the inertia tensor points in two
possible directions.
{}From the initial point, the sequence is propagated in both (opposing)
directions until termination, and the entire joined sequence is
what is used for statistical analysis, as long as the total number
of links is at least $N_{L,min}$.
Generally, sequences tended to be non-intersecting but in some cases they
oscillated between two points.  When this is detected, the
sequence is terminated.

\subsection{Computation of Statistics}

We developed three statistics to compute on a link sequence
which measure filamentarity or planarity in an easily interpretable way.
We call them planarity, curvature, and torsion.
These statistics are defined as angle deviations between
inertia ellipsoid axes for consecutive points along a link sequence; the exact
definitions are as follows:

\begin{itemize}

\item
{\it Planarity} ($\theta_P$)
is the angle difference between the minor axis of the inertia tensor for two
consecutive points.
The geometrical interpretation of planarity is as follows:
Given that filaments in large-scale structure often
occur at intersections of sheet-like structures, the minor axis of
the inertia tensor along the filament measures the strength of the
embedding sheet perpendicular to the filament;
hence a lower planarity angle indicates the presence of a
local sheet-like structure.

\item
{\it Curvature} ($\theta_C$) is defined as the angle difference between two
consecutive links.  Equivalently, it is the angle difference between the major
axis
of the inertia tensor for two consecutive points.
A sequence which is following a well-defined filament will have a low
angle difference between links;  hence a lower curvature angle indicates
greater filamentarity.

\item
{\it Torsion} ($\theta_T$) is
the angle difference between the plane defined by the
first two links and the third link.  Torsion measures the
strength of the embedding sheet parallel to the filament, a lower
torsion indicating a stronger planar structure present.

\end{itemize}

In all cases, {\it a lower value (angle difference) signifies more
structure present in the catalog}.  As an example, consider a
set of points distributed randomly throughout a long, thin cylinder.
A sequence will track the cylinder, and the angle deviation between
each successive link will be very small; hence curvature will show
a very low angle deviation.  Conversely, planarity and torsion will show
large angle deviations since there is no locally preferred plane
in a circular cylinder.
For a thin sheet, sequences will randomly walk throughout the sheet,
yielding a high curvature angle (indicating no filamentary structure), but
low planarity and torsion angles (indicating lots of planar structure).

In large-scale structure, filaments are often embedded within
sheets, and thus these statistics are expected to be
correlated.  Nevertheless it is useful
to consider each one separately.  A key difference between
the statistics is that each requires a different number of sequence points
to compute.  Planarity is the most {\it local} statistic, being computed
from only 2 link nodes, while curvature requires 3, and torsion
requires 4.  While planarity and torsion are in the ideal case purely
measures of planarity, torsion is more sensitive to the presence of
local filamentary structure since it measures angle
differences along the sequence rather than perpendicular to the sequence.

For each of those statistics, an average value
is found within a single sequence.  Then, for all the
sequences in that catalog, a median
value is found.  We will denote the resulting averaged-then-medianed
statistic by a bar,
as in $\bar\theta_C$.  This final median value is the value of that statistic
for the given catalog at the selected value of $R$.
Errors analysis is discussed in section 4.

\subsection{Visualization and Algorithm Testing}

We have attempted to construct an algorithm which will identify and
track filaments.  We tested the algorithm on artificially generated
point sets of lines and planes of varying thickness.  The results
conformed to qualitative expectations, that lines should show a
great deal of filamentarity and little planarity, and vice
versa for planes.  Also, the median angle deviations increased
with thickness, as expected.  Visualizations showed that link sequences
were tracking the structure as expected.

When we visualized the link sequences which were generated in an
actual CHDM simulation,  they
tended to lie preferentially in regions of structure,
but could not often be associated with visually
recognizable filaments.  They were also scattered throughout
the simulation volume.
This is because for the simulations we considered (which will be
described in the next section), nearly every galaxy that was tried
as a sequence starting point yielded a qualifying ($N_{P,min}\geq 4$)
sequence.  Thus the parameter set we have chosen does not sufficiently
restrict the generated sequences to lie directly along the
filaments that are detected by eye.  By imposing more severe
requirements for sequence qualification, one can tune the
algorithm to better recognize filamentary patterns.  However, this reduces
the number of qualifying sequences to a point where statistics
are poor, and hence it is not useful for performing
statistically significant comparisons.
Our conclusion is that this algorithm is not particularly
suited for pattern recognition, and
is better suited for statistical comparison of overall structural
properties of models.
The statistics we compute have simple interpretations, and the
results for various models are
consistent with the BHNPK visualizations;  however, this agreement is not
necessarily apparent from visualizations of individual link sequences.

Little effort went into developing analytical predictions for
expected values of $\bar\theta_C$, $\bar\theta_P$, and $\bar\theta_T$,
even in the case of a Poisson catalog.
This is due primarily to the fact that the algorithm was
successful in the test cases we considered, and thus a
complex and time-consuming analytical prediction was deemed
to be low priority.  Further numerical testing may also be done
by superimposing lines or sheets of varying strengths on a Poisson catalog,
and determining how effective the algorithm identifies structure.
We leave these endeavors to the future, and instead for now
concentrate on applications to the comparison of cosmological models.

\section{The Simulations and Data}

\def\kmsmpc{km/s$\cdot$Mpc}

\subsection{The Halo Catalogs}

The filament statistics were applied to the simulations
described in Klypin, Nolthenius \& Primack (1995; KNP95),
which are 100 Mpc$^3$ particle-mesh simulations
on a 512$^3$ force resolution grid.  All had $\Omega=1$ and $H_0 = 50$
km s$^{-1}$ Mpc$^{-1}$ (which will be assumed throughout).
A resolution element, or cell, is 195 kpc.
The CDM simulations had 256$^3$ particles, while the CHDM simulations
had 256$^3$ cold particles and $2 \times 256^3$ hot particles, giving
a cold particle mass of $2.9\times 10^9 M_{\odot}$ and $4.1\times 10^9
M_{\odot}$
for CHDM and CDM, respectively.  There were two simulations with pure
CDM, one with linear bias factor $b=1.0$ (CDM1) and one with $b=1.5$ (CDM1.5),
and two
CHDM simulations with $10\%$ baryons, $30\%$ in a single neutrino species and
the rest cold dark matter.

CDM1 and both CHDM simulations have linear bias factors which are
compatible with the COBE DMR results.
CHDM$_1$ and both CDM simulations were started with identical random number
sets describing the initial perturbation amplitudes.
It was found in NKP94, NKP95 and
KNP95 that Set 1 had, by chance, an unusually high power
($\sim \times 2$) on scales comparable to the box size.  However,
the CfA1 data appears to show similarly unusual power when
compared to the larger APM survey data (NKP95, Vogeley \etal 1992, Baugh and
Efstathiou 1993). CHDM$_2$ had a power spectrum more typical of a 100 Mpc box.
Thus CHDM$_1$ should be compared to the CDM simulations for
discrimination between models, while CHDM$_1$ can be compared to CHDM$_2$
to (conservatively) estimate cosmic variance.  Note that by using
identical random number set initial conditions, cosmic variance is
explicitly removed between the CDM1, CDM1.5 and CHDM$_1$ simulations.
Thus comparisons between these simulations reflect only differences
in the underlying physics of the models.
These four {\it halo catalogs} are summarized in Table 1.

Galaxies are identified initially as dark matter halos with ${{\delta\rho}/
{\rho}} > 30$ in 1-cell resolution elements (corresponding to about 4 cold
particles in a cell) which are local maxima in density. Masses were
computed in $3\times 3\times 3$ cells surrounding the maximum
(using 1-cell masses made no difference in the results), then
halos with $M > 7 \times 10^{11} M_\odot$
were broken up to address overmerging (NKP95).

We also tested filament statistics on catalogs in which we identified
galaxy halos as cells with ${{\delta\rho}/
{\rho}} > 80$.  These catalogs
gave basic results which were quite similar to the halo catalogs
described above, with a slight increase in Poisson errors due to
fewer numbers of halos.  While  the ${{\delta\rho}/
{\rho}} > 30$ catalogs have too many halos to be associated with visible
galaxies,
these catalogs still serve our purpose of testing whether these
statistics can quantify structure and discriminate between models in real
space.
Comparisons with real data must be done using simulated redshift-space
catalogs.

\subsection{The Sky Catalogs}

NKP94 and NKP95 describe the construction of the CfA1-like sky-projected
redshift catalogs from the simulations described in the previous section,
and the merged (to match simulation resolution) CfA1 catalog.
In order to distinguish these
catalogs which are designed to mimic many observational properties of the
CfA1 survey from the halo catalogs described above, we call the
CfA1-like sky-projected redshift catalogs the {\it sky catalogs}.
Several items in sky catalog construction which are
of relevance to filament statistics are:

(1) Six view points were chosen from within the CHDM$_1$
and CHDM$_2$ simulations satisfying the conditions that
the local density in redshift space ($V < 750$ km s$^{-1}$) is within
a factor of 1.5 of the merged CfA1 galaxy density, and the closest Virgo-sized
cluster is 20 Mpc away.  The CDM view points were required
to be on the halos nearest to the CHDM$_1$ view point coordinates,
and thus the corresponding sky catalogs, like the halo catalogs, differ
only because of their underlying model physics and not cosmic variance.

(2) To create a sky catalog of CfA1 size (12,000 km s$^{-1}$, 2.66 steradians),
the periodic halo catalogs were stacked, then cut to form the CfA1
survey geometry; hence structures appear typically $\sim 3-4$ times,
although distant galaxies are sampled sparsely.

(3) Each sky catalog was cut to CfA1 numbers before fitting a Schecter
luminosity
function (after monotonically assigning Schecter luminosities to mass).
The scatter in Schecter function parameters among the six view points
is thus convolved into the statistics.

\subsection{The Effect of Halo Breakup}

The most massive halos in the simulation should generally have more than one
individual
galaxy associated with them (Katz and White 1993, Gelb and Bertschinger 1994).
These ``overmerged'' halos were broken up as described
in NKP95 (it is the ``preferred method" set of catalogs that was used here).
Only 0.5\% of CHDM halos required breakup, raising the
number of halos with ${\delta \rho / \rho} > 30$ by $\sim$16\%.
CDM1.5 and CDM1 catalogs had higher fractions of massive overmerged halos;
1.3\% and 1.7\% respectively, raising their breakup halo populations by 35\%
and 56\%, respectively.
We expect the halo catalog results to be fairly insensitive to breakup since
they
probe scales $\sim$3 Mpc and up, much greater than the radius over which
fragments are distributed, which is typically $\lesssim 1$ Mpc.
Indeed we will show this to be the case in section 4.4.

Despite the larger scales investigated, sky catalogs will be more sensitive to
breakup.
This is because breakup takes a single massive halo and fragments it
into many closely-distributed objects,
many of which survive the magnitude limit.
When normalized to CfA1 number density, the net effect of breakup is to
weight the massive halos more strongly, giving the appearance on average of
moving galaxy halos into spherical groups (albeit with some ``finger of
God" elongation).  For a dense catalog, overdense
regions will be augmented at the expense of underdense regions, but
for sparse catalogs like CfA1, only the densest
clusters are augmented, at the expense of filamentary and planar structures.
Hence it turns out that breakup
tends to systematically {\it reduce} the amount of filamentary and
planar structure measured in sky catalogs.

\section{Application of Filament Statistics}

\subsection{Results for Halo Catalogs}

Filament statistics were applied to the above described
halo catalogs catalogs after breakup.
Figure~\ref{fig: dgfullstats}
shows the results for planarity $\bar\theta_P$, curvature $\bar\theta_C$,
and torsion $\bar\theta_T$ vs. $R$, where $R$ is in units of $\bar d$.
The statistics were computed for each $R$
from 1.0 to 2.5 in increments of 0.1, where discrimination levelled
off or began falling.
To estimate errors in the halo catalogs, each statistic was computed
over a random subset of the catalog.  The subset was taken to be as
many galaxies as necessary to generate 500 link sequences.  Even for
$R = 1.0$, this never required more than 535 galaxies;
at high $R$, hardly any galaxies generated sequences which
did not meet the $N_{L,min} = 4$ criterion.
The catalog was then resampled 10 times to obtain an error estimate.
Since there are more than 34,000 galaxies in each catalog, the data is
not oversampled.
At $R = 1.0$, there were on average 5--6 links per sequence; this number rose
steadily until $R \geq 1.6$, where sequences were was almost always
terminated due to the $N_{L,max} = {{\rm 100 Mpc}/ \bar d} \approx 34$
criterion.  The
average number of galaxies within a sphere of radius $R$ around a
given sequence point rose from
$\sim$10 at $R=1.0$ roughly linearly to $\sim$50 at $R=2.5$.

Figure~\ref{fig: dgfullstats} shows that all three statistics are
generally higher for the CDM simulations as compared
with the CHDM simulations, indicating that
CDM is less filamentary, has fewer sheet-like structures,
and has more (spherical) clustering than the CHDM simulations.
This is consistent with the BHNPK visualizations.  Thus
filament statistics do provide quantitative differentiation
between structure seen in the halo catalogs.

Note that all the statistics tend to fall with increasing $R$.
This reflects the fact that as the ratio of ${R/L}$ increases,
the greater overlap
between adjacent spherical windows generates stronger correlations
between adjacent inertia tensors, thereby reducing the angle deviations
between neighboring inertia ellipsoid axes.
There is an additional effect that is peculiar to catalogs possessing
inherent filamentary structure:
Consider a link sequence tracing a path defined by
points contained in a ``filamentary structure"
of radius $R_{cyl}$.  As we increase
${R/L}$ we see an increasingly more linear distribution of points in the
window, thus lowering the value of ${\bar\theta_C} \sim \frac{1}{2}
\arcsin(R_{cyl}/{R})$.  A similar argument holds for planarity
and torsion.  In reality, the galaxy distribution
is much more complex, but the basic result is that sampling large-scale
structure
gives ${\bar\theta_C}(R)$, ${\bar\theta_P(R)}$,
and ${\bar\theta_T(R)}$
falling at rates greater than in the Poisson case.

The large difference between simulations and the Poisson catalogs provides
a good indicator of how effectively structure
is identified by filament statistics.
Link sequences identify and follow structure in a Poisson catalog
by detecting chance alignments of halos which masquerade as contiguous
structure due to finite
numbers of halos in a given window.  We call this effect {\it structure
aliasing}.
Structure aliasing is primarily a low-galaxy-density phenomenon, and hence is
most significant at low $R$, where all sequences barely exceed $N_{L,min}$,
and each window barely has $N_{P,min}$ halos.  In this situation the
majority of sequences which qualify will be those lying along such
rare chance alignment of halos.  Increasing $N_{P,min}$ and $N_{L,min}$
reduces structure aliasing, but the corresponding reduction in qualifying
sequences increases shot noise significantly.  Instead, we simply
choose to be careful about our interpretations at low $R$.  For instance,
for $R \leq 1.3$ the Poisson catalog statistics rise with $R$,
indicating that aliased structure is significant here.
Structure aliasing occurs in the models as well, but
is less apparent because halos are correlated, yielding
more halos surrounding a given point than in the Poisson case.
Nevertheless the reduced discrimination for $R \leq 1.3$ is an
indication that aliased structure is of comparable strength
to real structure at these scales.

At low and high $R$ values, filament statistics discriminate between
the CDM models with different biases, as shown in Figure~\ref{fig:
dgfullstats}.
At $R \leq 1.3$ CDM1.5 aliases structure more effectively than CDM1
since it is more diffuse (more Poisson-like),
while at larger scales ($\ga 7$ Mpc) the enhanced clustering of CDM1.5
(see BHNPK) tends to
trap sequences in spherical clumps more effectively than CDM1, giving
higher values.
Identification of these effects over a $R$ of
1.0--2.5 (roughly 3.0--7.5 Mpc) indicates the high sensitivity of these
statistics to the presence of structure.

The two CHDM simulation results are within each other's error bars on the
scales investigated.
Hence for these statistics, cosmic variance between CHDM$_1$ and CHDM$_2$
may be comparable to resampling errors in the halo catalogs.

As in Hellinger \etal (1995), here we have introduced a
set of metastatistics to compare statistics
and assess the effectiveness of our analysis.
Discrimination between models for a given statistic $\theta$ can be measured by
the {\it signal strength} $S^\theta_{\rm res}$ between catalogs:

\begin{equation}
S^\theta_{\rm res}(1,2)={{| \theta_1-\theta_2 |}\over
{\sqrt{\sigma_{\theta_1}^2+\sigma_{\theta_2}^2}}}
\label{eq: dgsignal}
\end{equation}

\noindent
where $\theta_1$ and $\theta_2$ are values of statistic $\theta$
for catalogs 1 and 2, respectively,
and $\sigma_\theta$ is the resampling error for that statistic and catalog.
The subscript ``res" denotes that the units of $S^\theta_{\rm res}$ are
1-sigma resampling errors.
To compare CHDM to CDM at COBE normalization while minimizing noise due to
cosmic variance, we compare CHDM$_1$ to CDM1.
Figure~\ref{fig: dgsig}(a) shows $S^\theta_{\rm res}$(CDM1,CHDM$_1$) for the
halo catalogs
for $\theta = \bar\theta_P, \bar\theta_C, \bar\theta_T$.
Planarity shows the highest signal, but at low $R$ this is spurious since
planarity is the most local statistic (requiring only two links for
computation) and hence is most sensitive to structure aliasing.
The planarity signal between CHDM$_1$ and CDM1 is high at low $R$, but between
CHDM$_1$ and CDM1.5 it is much lower (as seen in Figure~\ref{fig:
dgfullstats}),
indicating that $\bar\theta_P$ is the least robust of
the statistics against structure aliasing.
For $R \geq 1.5$, where structure aliasing is unimportant,
all statistics show comparable discrimination of
$S^\theta_{\rm res}$(CDM1,CHDM$_1$)$\ga 4\sigma$.
Thus filament statistics are fairly discriminatory for the halo catalogs.
They are also quite robust; their robustness against halo breakup
will be formally investigated later.

\subsection{The Effect of Redshift Distortion}

Redshift distortion is potentially a major concern for filament statistics
when applied to the sky catalogs.
Naively, one might assume that filament statistics do not discriminate effects
due
to internal velocities in clusters (``fingers of God") from genuine linear
structures.  This is in general not the case, since link sequences contain
directional information.
Fingers of God are elongated only in the line-of-sight direction \^{\bf r}
in redshift space,
whereas in general a filament in real space will not be aligned with \^{\bf r}.
Since the largest fingers of God occur at the intersection of real filaments,
another effect of redshift space distortion is to misguide sequences and
increase the angle deviation of a sequence passing through a cluster.
The net effect is seen as a decrease in the amount of detected structure
in these simulations.

To test the effect of redshift distortion we applied filament statistics to
halo catalogs with overdensity ${\delta \rho}/{\rho} > 80$
as we scaled the halo peculiar velocities from 0 to a velocity factor $F_V = 5$
times their actual
values.  These catalogs were then mock-observed (no magnitude limit)
from the
origin, including the effect of redshift distortion, at each $F_V$.
$F_V = 0$ represents real space, $F_V = 1$ represents normal redshift space,
and $F_V$ is increased until fingers of God become large enough to dominate
structure identification.  Note that $L=\bar{d} \sim 4-5$ Mpc in these
catalogs,
and we have taken $R=1.5L$.  Error bars are obtained in the same way as in
the halo catalogs, by taking 500 galaxy random subsamples and resampling 10
times.

Figure~\ref{fig:reddist}(a) shows
the effect of the transformation from real to redshift space upon filament
statistics for the ${\delta \rho}/{\rho} > 80$ halo catalogs.
At a $F_V = 1$, we see little change in the discrimination
between the models from $F_V = 0$, perhaps even an increase
in discrimination.
Even with velocities scaled to 3 times their actual values the models are still
well discriminated.  At $F_V \geq 4$, fingers of God begin to dominate,
as evidenced by the flattening of statistical values vs. $F_V$,
and the increased resampling error.
This test was also run on no-breakup versions of the ${\delta \rho}/{\rho} >
80$
halo catalogs,
and it was found that the interpretations are virtually independent of
breakup in both real and redshift space.

This test indicates that fingers of God are not a dominant contributor to the
measured structure when applied to halo catalogs with overdensity ${\delta
\rho}/{\rho} > 80$.
The primary effect of redshift distortion is to
decrease strength of structure.  Trends vs. $F_V$ appear not to be highly
model-dependent.

To test how density affects redshift distortion, we also ran the
velocity scaling test for the ${\delta \rho}/{\rho} > 30$ halo catalogs,
and to catalogs with an even
greater overdensity, ${\delta \rho}/{\rho} > 120$.
The ${\delta \rho}/{\rho} > 30$ halo catalog results confirmed
that the primary effect of redshift distortion
is to decrease the amount of structure detected;
redshift distortion had even less effect on model discrimination than
in the ${\delta \rho}/{\rho} > 80$
halo catalogs, and little evidence of distortion was
seen even out to $F_V = 5$.

Figure~\ref{fig:reddist}(b) shows the more interesting results for the
halo catalogs with overdensity ${\delta \rho}/{\rho} > 120$.
Here, redshift distortion
increased resampling errors (thereby degrading the discrimination between
models)
somewhat even at $F_V = 2$, and progressively more severely
at higher $F_V$.  This indicates that
higher overdensity cuts make redshift distortion a more severe problem.
This is expected, since high density regions in general have higher
peculiar velocities.
Still, at $F_V = 1$ ({\it i.e.,} ordinary redshift space)
the discrimination is actually increased
over real space.  This is because
the length and abundance of fingers of God reflect clustering
properties, which show true differences between models (NKP94, NKP95).
The characteristically longer fingers of God in CDM (due to higher
halo peculiar velocities) adds extra dispersion in the link directions
which adds to the lesser real space filamentarity in CDM.  The result
is an amplification of the differences between CHDM and CDM for all of
the filament statistics we compute. Curvature and torsion are
amplified more than planarity because
redshift distortion tends to produce more artificial filaments
than artificial sheets, and both curvature and torsion measure angle
differences along the filament rather than perpendicular to the filament.
In sky catalogs, this amplification is more than counteracted by the
increase in statistical error due to lower sample size.

\subsection{Results For Sky Catalogs}

We present results of filament statistics applied to the
sky catalogs after halo breakup in Figure~\ref{fig: dgskystats}.
Every galaxy in each sky catalog was tried
as a possible sequence starting point.  For each catalog, at $R = 1.0$,
around 500 of the $\approx$2360 galaxies typically generated sequences with
number of links exceeding $N_{L,min}=4$.  This number rose roughly linearly
until
$R=2.5$, where $\sim$2200 galaxies qualified, on average,
in each catalog.  There were systematic differences between the catalogs
as well, with CHDM$_2$ showing the largest number of accepted sequences,
about $5-10\%$ more than the CDM models.  CHDM$_1$ showed the
lowest number, consistently slightly below the CDM models.
At $R = 1.0$, there were on average about 5 links per sequence;
this number rose fairly linearly with $R$, such that at $R=2.5$, there
were around 20 links per sequence.  The average number of galaxies within
a sphere of radius $R$ around a given sequence point rose from 8--10
at $R=1.0$ roughly linearly to 25--30 at $R=2.5$.

The error estimate for each statistic in sky catalogs was
determined from {\it sky variance}, by computing the statistic at
each of six vantage points, and getting an average value and standard
deviation for that statistic.
Since our box is relatively small, different viewpoints are still
seeing many of the same structures, although with differing depth.
Sky variance is therefore expected to underestimate true cosmic
variance, perhaps significantly in some cases.
With current simulations, a better method of estimating cosmic variance is
by comparing CHDM$_1$ and CHDM$_2$.  NKP94, NKP95 and KNP95 estimate the high
power
in CHDM$_1$/CDM1/CDM1.5 would be expected $\sim 10\%$ of the time,
translating to a $\sim 1.7\sigma$ deviation from norm, while CHDM$_2$
was found to be quite typical.  Thus CHDM$_1$ vs. CHDM$_2$ may be
taken as a conservative estimate of $1\sigma$ cosmic variance.

Figure~\ref{fig: dgskystats}
shows the results of filament statistics applied to the sky catalogs
after halo breakup.
CHDM$_1$ still shows more structure than either CDM, but CDM1 and CDM1.5
are not discriminated.
The interesting new feature is that the two different sets of initial
conditions are now discriminated, with
CHDM$_2$ values being lower than CHDM$_1$ at low $R$, and higher than
CHDM$_1$ at high $R$.  Statistics on CDM1 and CDM1.5 show a similar
behavior to CHDM$_1$, indicating that the reason for this difference
is because the scales are large enough
(median $\bar d \sim 15$ Mpc) for the anomalously high large-scale power in
initial conditions set 1 to become significant.
The extra large scale power in set 1, combined with
the artificial replication of structure in the construction
of the sky catalogs at 100 to $100\sqrt{3}$ Mpc intervals,
gives more structure at large scales in set 1 models than in CHDM$_2$.
Simulations with sufficient volume should not suffer from this problem.
The CfA1 catalog follows CHDM$_2$
more closely than the other simulations over $R\approx 1.2-2.0$, where
discrimination from the Poisson catalog is best and the statistics are
therefore
most reliable.

Visualization showed that link sequences were distributed throughout
the sky catalog volume, with very few lying in the foreground, $r\lesssim 20$
Mpc.  Recall that
$\bar d (r)$ is small at low $r$, and the Virgo Cluster,
being nearby,
contributes hardly any sequences even though it gives a large
finger of God.  At small $R$, sequences tended to be shorter and
terminate within the catalog volume, while at large $R$ they tended
to terminate once they exceeded the catalog boundary and found no
nearby galaxies.  Also, at large $R$ the sequences tended to
be preferentially radially directed, because the spheres of
radius $R$ tended to extend beyond the catalog volume, and hence
the entire catalog contributed as a single radial filamentary structure.
This was clearly evident for $R\ga 2.0$, indicating that results
for these $R$ values are of dubious validity.

In the halo catalog comparisons we estimated Poisson errors for the statistics
and left cosmic variance implicit in the comparison of CHDM$_1$ and CHDM$_2$;
for the sky catalogs we must consider the total cosmic variance which includes
variations due to different realizations of the models as well as the choice of
view points in redshift space.  CHDM$_1$ shows more structure than CHDM$_2$
by $\ga 2\sigma$ (where $\sigma$ are sky variance errors) for all $R \geq 1.5$.
We interpret this to mean that sky variance is an inadequate estimate of
cosmic variance, which is clearly too large to discrimate
between CDM and CHDM for these values of $R$.
This effectively restricts the discriminatory ranges of $R$ to $R \approx
1.2-1.3$,
and suggests that
the sensitivity to cosmic variance is approaching the sensitivity to
model parameters.  Hence we should be increasingly concerned with
more closely comparable local environments as well as more realistic models.

Figure~\ref{fig: dgsig}(b) shows $S^\theta_{\rm sv}$(CDM1,CHDM$_1$) for the sky
catalogs,
for $\theta = \bar\theta_P, \bar\theta_C, \bar\theta_T$.
This comparison emphasizes the signal strength with respect to sky variance
(denoted by subscript ``sv"),
as we have intentionally reduced
cosmic variance by comparing simulations started from the same initial random
numbers.
Discrimination between CHDM$_1$ and CDM1 is strongest in curvature and
torsion are, while planarity shows a reduced signal.  Planarity is weaker
because it is not as significantly amplified by redshift distortion
as curvature and torsion, as was shown in
section 4.2 (see Figure~\ref{fig:reddist}(b)).
Recall that CDM shows stronger clustering, which leads to greater
redshift distortion, which in turn leads to less structure being
detected by these statistics.  This accentuates the real space
differences between CDM and CHDM.
Thus filament statistics convolve
information from clustering properties of models when applied
to CfA1-like catalogs;
sensitivity to clustering will reduce in denser surveys.

To test sensitivity to shot noise, filament statistics were applied
to full-sky versions of the sky catalogs, which covered 10.384 steradians
and contained $\approx 9200$ galaxies (nearly 4 times the CfA1-like sky
catalogs).
Since the 2.66 sr catalogs and the 10.384 sr catalogs are derived
from the same simulation data set, we are
still sampling from the same distribution of cluster sizes and shapes.
The resulting signal increased by a factor of $\sim 2$ (for $R \geq 1.3$) as
expected from sample size statistics.
The degradation of the signal from the halo catalogs to the sky catalogs
($S^\theta_{\rm sv} \sim 2.5$ for $R > 1.6)$ is thus primarily due to
sparseness.

The results before breakup are not shown, but as described in section 3.3
the catalogs before breakup show slightly more structure than after breakup.
It turns out this effect
represents a $\lesssim 1^\circ$ increase in each statistic for the
sky catalogs, which is
comparable to sky variance errors.  There is little difference in
$S^\theta_{\rm sv}$(CDM1,CHDM$_1$) for no-breakup sky catalogs.  Robustness
against halo breakup
will be formally investigated in the next section.

The statistics were also applied to 80 Mpc volume-limited versions of the
CfA1-like sky
catalogs, and left typically 400-500 galaxies in each sky
catalog.  The statistics showed very large shot-noise
scatter, and gave no significant discrimination between models.
Volume limiting certainly yields more interpretable statistics,
but for CfA1 and our similar-size simulation sky catalogs,
there are simply too few galaxies.

\subsection{Robustness Against Halo Breakup}

Galaxy identification represents the single biggest uncertainty in catalog
construction, both in halo catalogs and in redshift space.
To investigate robustness against halo breakup and halo identification
we take the halo catalogs and sky catalogs before and after
breakup, find the change in the value of a given statistic
for each $R$,
and average the difference over all realizations (CDM1, CDM1.5, CHDM$_1$ and
CHDM$_2$).  Formally, we define the {\it galaxy identification uncertainty
factor}

\begin{equation}
F_{\rm id}(\theta)=\left\langle{{| \theta_{\rm bu}-\theta_{\rm nobu} |}\over
{\sqrt{\sigma^2_{\theta_{\rm bu}}+\sigma^2_{\theta_{\rm
nobu}}}}}\right\rangle_{{\rm realizations}}
\label{eq: dgrob}
\end{equation}

\noindent
where $\theta$ represents the value of the statistic in question, ``bu" and
``nobu"
refer to breakup and no-breakup catalogs,
and $\sigma_{\theta}$ represents the error on that statistic, which
are from resampling or sky variance.
$F_{\rm id}(\theta) << 1$ indicates a statistic which is robust against
galaxy identification and halo breakup.

We then compute the {\it combined signal strength}
by combining the resampling error (for halo catalogs) or sky variance
error (for sky catalogs) in quadrature with galaxy identification
uncertainty.  For halo catalogs,

\begin{equation}
S^\theta_{\rm res+id}(1,2)={{S_{\rm res}^\theta(1,2)}\over {\sqrt{1 + F_{\rm
id}^2(\theta)}}}
\label{eq: dgsrob}
\end{equation}

\noindent
where definitions are as in equation~\ref{eq: dgsignal}.
For sky catalogs, the subscript ``res" is replaced by ``sv", since
sky variance is the relevant error measure.

For the halo catalogs, we compute
$S^\theta_{\rm res+id}({\rm CHDM}_1,{\rm CDM}_1)$ for each statistic
for $R = 1.0-2.0$.  The results are plotted in Figure~\ref{fig: dgrob}(a).
The plot only goes up to $R=2.0$, not 2.5 as before, since
halo breakup is more significant at smaller scales.
Curvature and torsion show little degradation of signal as for $R \geq 1.5$ as
compared to
Figure~\ref{fig: dgsig}(a), indicating that these statistics
are quite robust against halo breakup in the range of $R$ where
structure aliasing is unimportant.
Planarity is less robust, since it is the most sensitive to structure aliasing.
In summary, all statistics show $\ga 4\sigma$ discrimination between models
regardless of halo breakup.

For the sky catalogs, the robustness against halo identification is not as
compelling, as
seen by the significantly lower values of $S^\theta_{\rm sv+id}$ shown in
Figure~\ref{fig: dgrob}(b)
as compared to $S^\theta_{\rm sv}$ shown in Figure~\ref{fig: dgsig}(b).
Now, only $1-2\sigma$ discrimination is seen, and only at specific
values of $R$.
This is a result of the $\lesssim 1^\circ$ increase in each statistic
from increased clustering due to breakup, as described in section 4.4.
This reflects
the seriousness of the overmerging problem for observational catalog
comparisons when using these statistics.  While we have devised a statistic
that is very robust with
respect to errors in the locations of galaxies as evidenced by the robustness
of the halo catalogs, robustness with respect to
the identification of galaxies is a much
more pernicious problem for the sky catalogs.

\subsection{Reduced Filament Statistics}

Reduced filament statistics were defined earlier as the value of each
statistic for a given catalog at a value of $R$ where the
Poisson catalog had its maximum discrimination from the survey data.
This definition is motivated by considering the Poisson catalog as
the ``noise" level for these statistics, the survey as the ``signal",
and identifying $R=R_{opt}$ where the signal-to-noise ratio is
maximized.

For the halo catalogs, this definition yields an $R_{opt}$ which is not
unique, since the signal-to-noise ratio is very large for all statistics and
catalogs for all $R \geq 1.5$.  Thus reduced filament statistics
yield no more information regarding model discrimination
than $S^\theta_{\rm res}$ for the halo catalogs.

For the sky catalogs, the results are more interesting.  The
signal-to-noise ratio hits a maximum for {\it all} statistics at
$R\approx 1.3$, and falls rapidly for smaller
or larger $R$ values,  suggesting $R_{opt}=1.3$.
The optimization occurs because at smaller scales the statistics are
dominated by structure aliasing, and at larger scales the
sampling windows for inertia tensor computation
will more often extend outside the catalog volume,
thereby confusing the axis determination and increasing noise.
The values of $\bar\theta_P$, $\bar\theta_C$, and
$\bar\theta_T$ applied to the sky catalogs at $R_{opt}=1.3$ are
shown in Table 2.  Also shown is the deviation from
the merged CfA1 catalog in units of the error for that statistic,
$\Delta_{\theta} = {{(\bar\theta - \bar\theta_{\rm CfA1})}/{\sigma_\theta}}$.

{}From Table 2 it is clear that the simulation which agrees
best with CfA1 for all statistics is CHDM$_2$.  The CDM models are ruled out
at a $3.1\sigma$ level from $\bar\theta_T$, and at a $2.4\sigma$ level
from $\bar\theta_C$.
There is hope that cosmic variance does not significantly degrade these
conclusions,
since CHDM$_1$ and CHDM$_2$ lie within $1\sigma$ of each other.

Analysis of the no-breakup sky catalogs shows
that both CDM1 and CHDM$_2$ are marginally consistent ({\it i.e.} CfA1
lies directly in between), and the other models are ruled out at more
than a $2\sigma$ level.  Thus reduced filament statistics, like full filament
statistics,
are not very robust with respect to halo breakup.

\section{Conclusions, Future Work, and Connection With Other Statistics}

Filament statistics applied to the halo catalogs are sensitive and robust
diagnostics
of large scale structure that effectively discriminate CDM models from
CHDM models.  The curvature statistic shows a
robust discrimination of $4\sigma$
between CDM and CHDM models.  Resampling variance is low, and
the result is insensitive to details of galaxy identification.  The
signal-to-noise ratio
between any model and the Poisson catalog is very large for
all $R \geq 1.5\bar{d}$, where $R$ is the window radius.

Comparison with CfA1 data is done by creating a sample of CfA1-like
redshift catalogs from each of the simulations,
and comparing these ``sky catalogs" directly to the CfA1 survey.
When one views the filament statistics results for sky catalogs over all values
of
$R$, it is clear that no model tested is completely
consistent with CfA1 data.
However, both the full and the reduced filament statistics
show that, at face value, the CHDM simulation with the more typical initial
conditions provides the best fit to CfA1 data.
For the reduced filament statistics, conservative estimates are that
the CDM model with $b=1.0$ is inconsistent with CfA1 at the $2\sigma$ level,
the CDM model with $b=1.5$ is inconsistent at the $3\sigma$ level,
and the CHDM model with the less typical initial conditions
is barely inconsistent at the $1\sigma$ level.  However, these results are
not robust with respect to details of galaxy identification.

The success of filament statistics for the halo catalogs indicates that
larger, denser redshift surveys coupled with larger simulations
will provide a significant
increase in the robustness and discriminatory power of these statistics
versus real survey data.  Denser surveys will lower sensitivity to
redshift distortion and lower shot noise in filament statistics.
We are looking forward to applying these statistics to other redshift
catalogs such as SSRS2, CfA2, and
the Sloan Digital Sky Survey, and comparing these data sets to the
latest in the rapidly-progressing field of cosmological simulations.

In a broader context, we view this work as illustrative
of a methodology for constructing new statistics to analyze spatial data.
We described the creation of link sequences, which produces data subsamples
extracted and organized to amplify properties of interest
in the underlying data set.  We emphasize that this is especially important
in analyzing nonlinear gravitational structures due to their complex
geometries and topologies.
A key distinction for filament statistics is that the new point set is
{\it guided} by the distribution of galaxies, not {\it bound} by it
(as in Delaunay or Voronoi tessellations, see e.g. van de Weygaert 1991)
and thus is more likely to be robust against variations in the galaxy
locations and halo breakup, though as we have seen, robustness
against galaxy identification is a separate issue.
The link sequence approach was conceived of as an intuitive means
of simplifying the complex topology of the galaxy point set while
preserving the sense of approximate connectivity of
its large-scale isodensity surfaces (which the eye might recognize as
``filamentarity'').

While we have not developed an analytical prediction for the values
of filament statistics, we have tested the algorithm by
visualizing the resulting link sequences from artificial
configurations of points as well as from simulation data sets.  The
algorithm does not perform well as a method for
identifying individual filaments within a simulation, but
by taking large samples of sequences one can obtain
statistically significant results which are consistent with
BHNPK visualizations of the simulations.

There are many other statistics one can compute on the link
sequences when viewed as spatial trajectories;  we have only
considered their most elementary discrete geometric properties.
For example, shape statistical filters (Hellinger \etal 1995)
can characterize local structures much more effectively than inertia axes.
With the rapid progress of computational technology and observational
data, filament statistics and other geometric network statistics
look to form a new and independent
class of statistics against which cosmological models may be tested.

\pagebreak
\section{Acknowledgements}
We acknowledge grants of computer resources by IBM, NCSA, SCIPP,
UCO/Lick Observatory, and UCSC Computer Engineering.
AK, JRP, and DH acknowledge support from NSF grants and
DH also acknowledges support from a DOE grant.
RD acknowledges support from the NSF GC3 grant and Lars Hernquist.
The simulations were run on the Convex C-3880
at the NCSA, Champaign-Urbana, IL.

\def\apj{ApJ}
\def\apjs{ApJ Supp}
\def\apjl{ApJ Lett}
\def\mnras{M.N.R.A.S.}
\def\aa{Astron. \& Astrophys.}
\def\aj{AJ}


\pagestyle{empty}

\newpage
\section*{Tables}

\bigskip
$$\vbox{\hsize=6.0in
\halign {\tabskip=0.9em #\hfil & \hfil #\hfil& \hfil #\hfil& \hfil #\hfil&
\hfil #\hfil& \hfil #\hfil& \hfil #\hfil\cr
{Model} & {Components} &  {Bias} & {Init.Cond.} & {No. of Gals.} & {$\bar
d$(Mpc)} \cr
CDM1 & $\Omega_{c}=1.0$ & $b=1.0^a$ & Set 1 & 58,121(37,164) & 2.58(3.00) \cr
CDM1.5 & $\Omega_{c}=1.0$ & $b=1.5\;$ & Set 1 & 61,690(45,592) & 2.53(2.80) \cr
CHDM$_1$ & $\Omega_{c}=0.6 \;\; \Omega_{\nu}=0.3 \;\; \Omega_{b}=0.1$ &
$b=1.5^a$ &Set 1& 34,000(29,151) & 3.09(3.25) \cr
CHDM$_2$ & $\Omega_{c}=0.6 \;\; \Omega_{\nu}=0.3 \;\; \Omega_{b}=0.1$ &
$b=1.5^a$ &Set 2& 34,554(29,765) & 3.07(3.23) \cr
\noalign{\hsize=6.0in
\noindent Table 1: Table of {\it halo catalogs} (~\cite{KNP95},~\cite{NKP95}).
The number of galaxies and
mean interparticle spacing $\bar d$ computed before halo breakup are
indicated in parentheses. \\
$^a$COBE-compatible bias.}}}$$

\bigskip

$$\vbox{\hsize=6.0in
\halign {\tabskip=2em plus 2em#\hfil & \hfil #\hfil& \hfil #\hfil& \hfil
#\hfil&
\hfil #\hfil& \hfil #\hfil& \hfil #\hfil\cr
{Catalog} & {$\bar\theta_P$} &  {$\bar\theta_C$} & {$\bar\theta_T$} &
{$\Delta_P$} & {$\Delta_C$} & {$\Delta_T$} \cr
CDM1 & 42.36$\pm$0.99 & 43.97$\pm$0.56 & 21.73$\pm$0.41
&1.6$\sigma$&2.4$\sigma$&3.1$\sigma$ \cr
CDM1.5 & 42.05$\pm$0.66 & 44.66$\pm$0.35 & 22.00$\pm$0.46
&1.9$\sigma$&5.8$\sigma$&3.4$\sigma$ \cr
CHDM$_1$ & 40.25$\pm$0.85 & 41.46$\pm$0.82 & 19.56$\pm$0.86
&-0.6$\sigma$&-1.4$\sigma$&-1.0$\sigma$ \cr
CHDM$_2$ & 40.71$\pm$0.43 & 42.02$\pm$0.70 & 20.11$\pm$0.43
&-0.1$\sigma$&-0.9$\sigma$&-0.8$\sigma$ \cr
CfA1 & 40.77 & 42.64 & 20.45 & - & - & - \cr
\noalign{\hsize=6.0in
\noindent Table 2: Table of reduced filament statistics
($R_{opt}=1.3$) for sky catalogs after breakup.
For each model, the value of each statistic is shown with sky variance errors,
and the deviation from the merged CfA1 catalog is shown in units of that
statistic's error ($\Delta_\theta$).
Models for which $| \Delta_\theta | \lesssim 1$ are
compatible with CfA1; clearly the CHDM simulations are more compatible.}}}$$


\newpage
\section*{Captions}

\begin{figure}[hp]
\plottwo{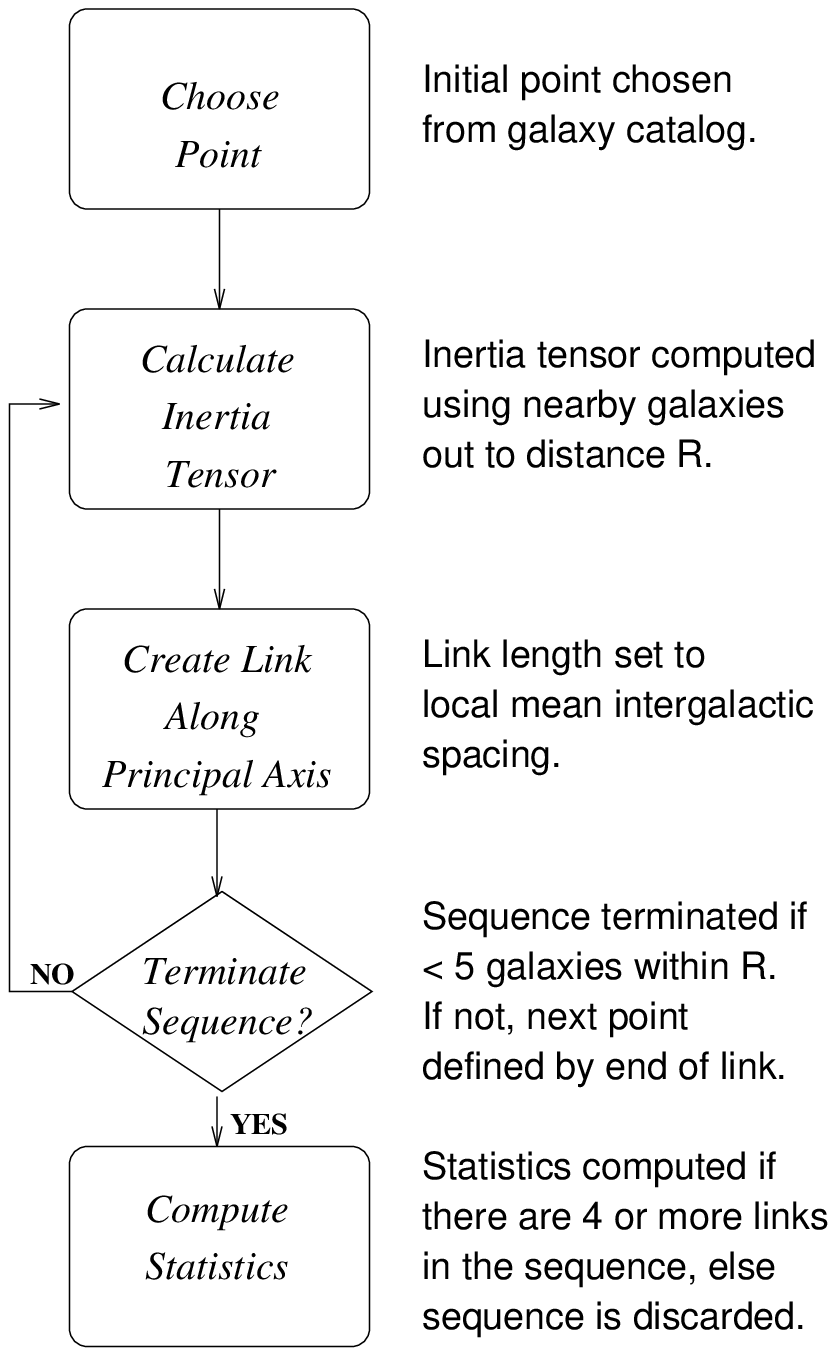}{blank.ps}
\caption{ }
\label{fig: dg_flowchart_filament}
\end{figure}

\bigskip
Figure~\ref{fig: dg_flowchart_filament}.
Link sequence generation computational flowchart.  Here, $R$ is in units
of the mean intergalactic spacing $\bar d$.  For galaxies in
redshift space, $\bar d$ is a function of the Hubble distance
$r = v/H_0$, where $v$ is the radial velocity of the galaxy.
{}From the initial galaxy, sequences are propagated in both (opposing)
directions
along the major axis until termination; if the combined number of links
is 4 or more, the entire (combined) sequence ``qualifies" for computation; else
it is discarded.

\begin{figure}[hp]
\plotone{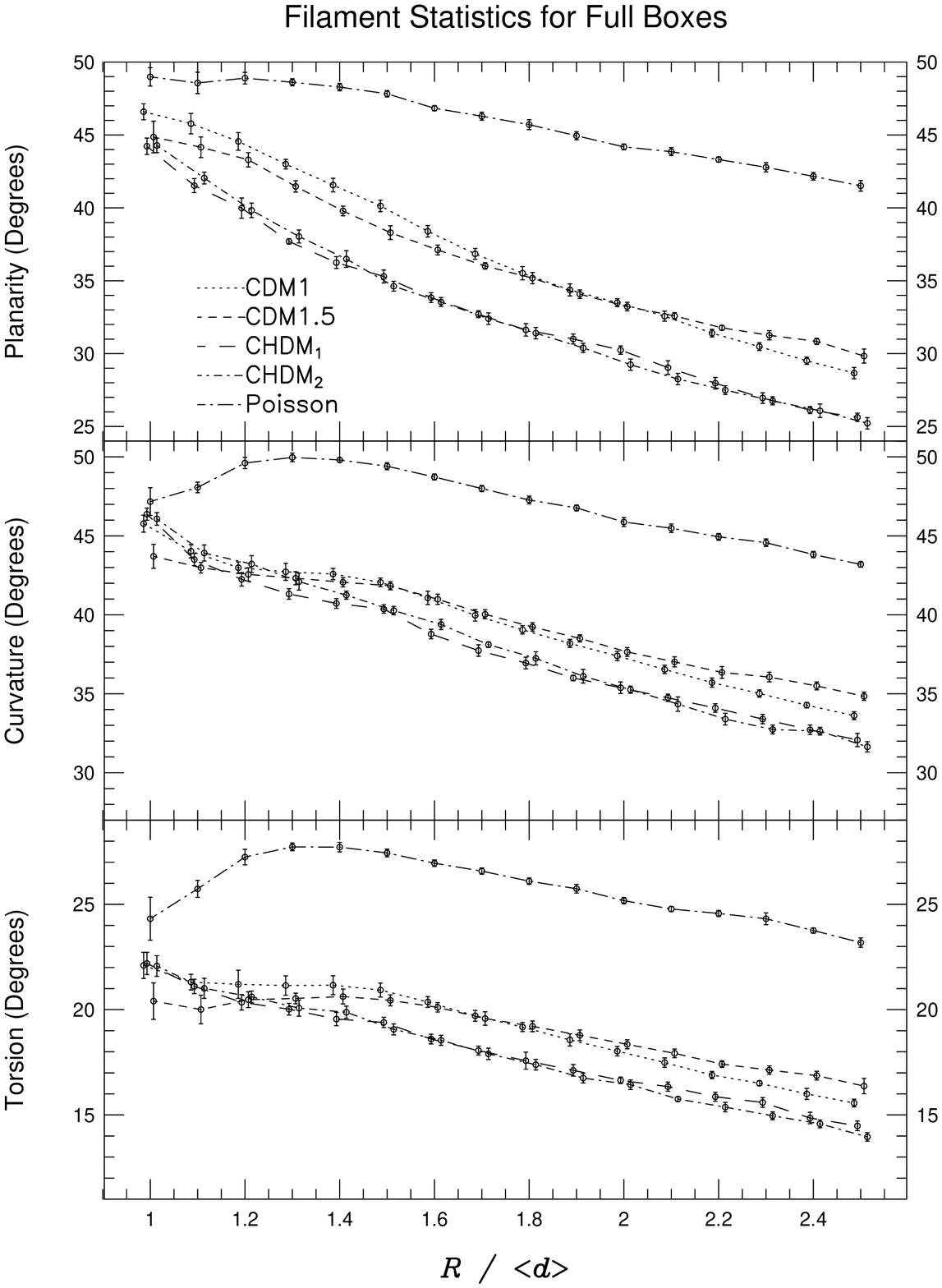}
\caption{ }
\label{fig: dgfullstats}
\end{figure}

\bigskip
Figure~\ref{fig: dgfullstats}.
{Filament statistics (planarity $\bar\theta_P$, curvature $\bar\theta_C$, and
torsion $\bar\theta_T$)
for the {\it halo catalogs} versus $R$ in units of
{\em $\bar d$}, the mean interparticle spacing, for
$L=\bar d$.  This plot
shows that filament statistics clearly and consistently
discriminate between CDM and CHDM for $R\ga 1.4$.
Even the different CDM biases are discriminated at certain $R$ values.
Cosmic variance estimated by the difference between
CHDM$_1$ and CHDM$_2$ is generally smaller than resampling error.
The Poisson catalog is well discriminated from any model
for $R\geq 1.4$.
Note: Values for different models are slightly offset in $R$ to improve
visibility.}

\begin{figure}[hp]
\plotone{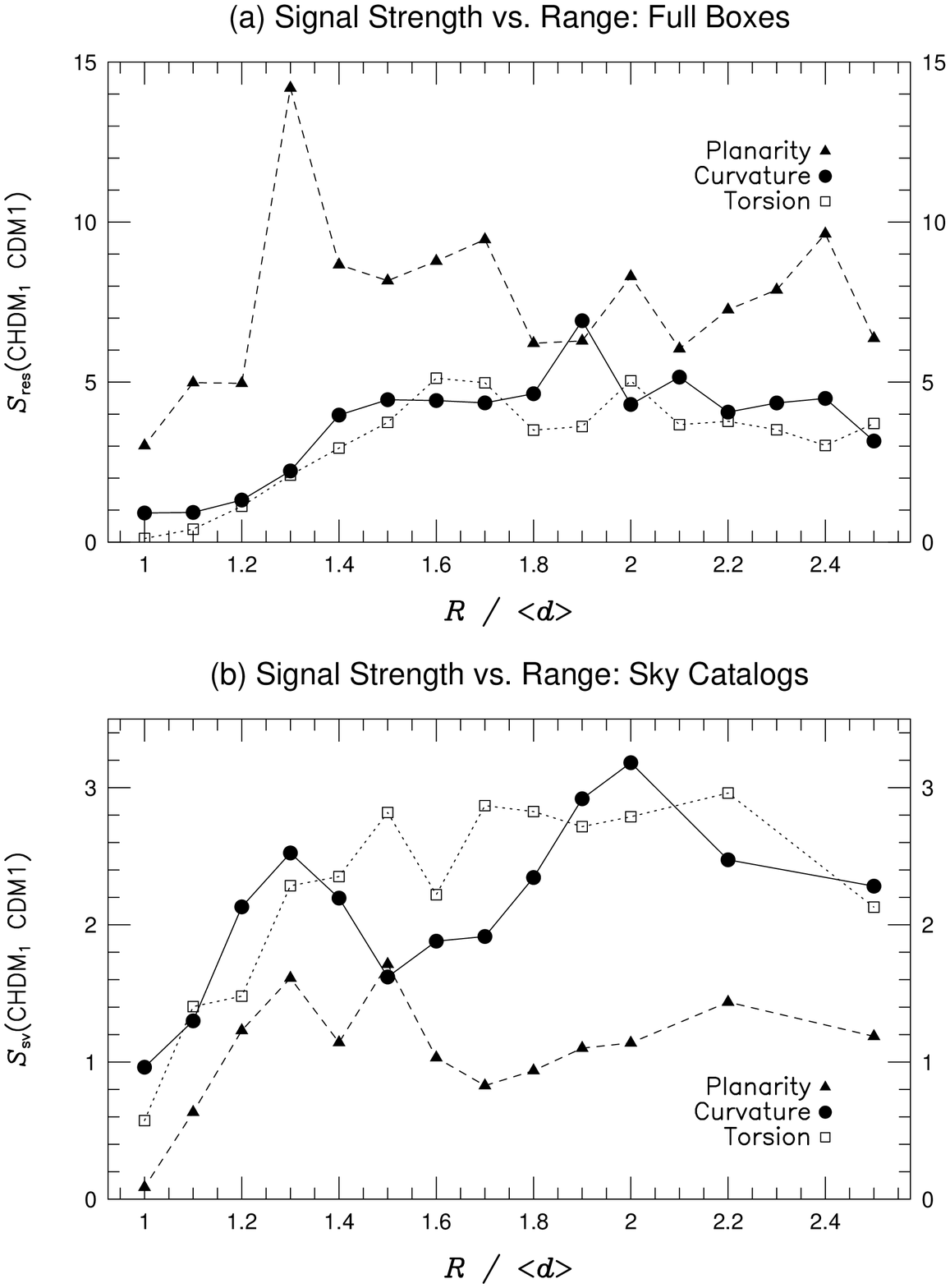}
\caption{ }
\label{fig: dgsig}
\end{figure}

\bigskip
Figure~\ref{fig: dgsig}.
{(a) Signal strengths $S^\theta_{\rm res}$(CHDM$_1$,CDM1), as defined in
equation (~\ref{eq: dgsignal}),
for  $\bar\theta_P$, $\bar\theta_C$, and $\bar\theta_T$ applied to
the halo catalogs.  All statistics discriminate fairly well for $R\ga 1.5$.
The high $\bar\theta_P$ signal for $R\lesssim 1.5$ is spurious,
owing to structure aliasing at small scales.
(b) Signal strengths $S^\theta_{\rm sv}$(CHDM$_1$,CDM1) for the
sky catalogs.  Curvature and torsion are better discriminators than
planarity, and stay around 2$\sigma$ for $R\geq 1.3$.}

\begin{figure}[hp]
\plotone{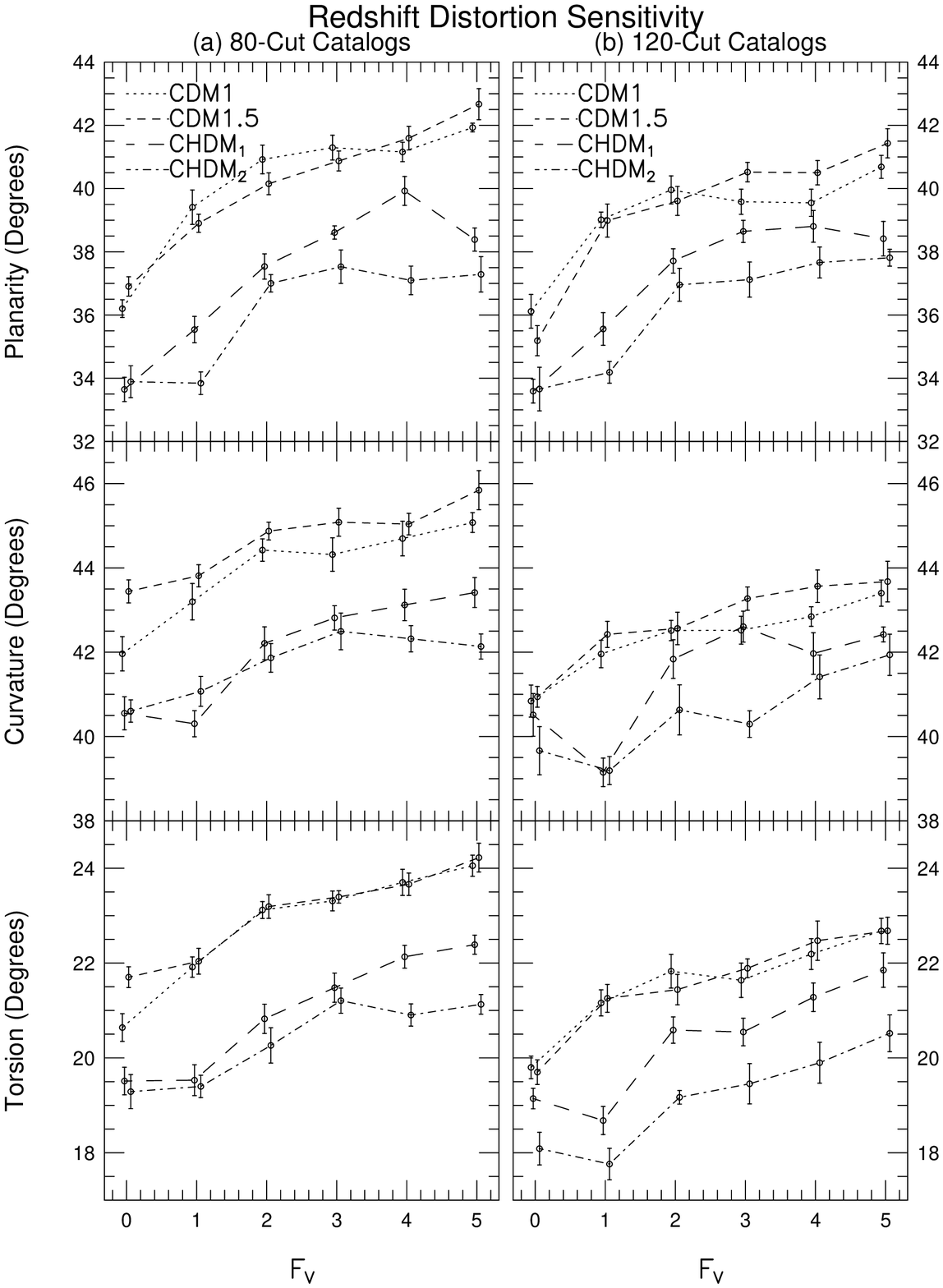}
\caption{ }
\label{fig:reddist}
\end{figure}

\bigskip
Figure~\ref{fig:reddist}.
{Filament statistics $\bar\theta_P, \bar\theta_C, \bar\theta_T$ applied
to mock-observed (a) ${\delta\rho/
\rho} > 80$ and (b) $\delta\rho / \rho > 120$ halo catalogs
with velocities scaled from velocity factor $F_V=0$ (real
space) to $F_V=5$ times their actual value, with $R=1.5$.
Going from real
space  ($F_V=0$) to ordinary redshift space ($F_V=1$) decreases
the amount of structure detected, but actually increases discrimination
between models.  In the $\delta\rho / \rho > 80$ halo catalogs, only for  $F_V
\ga 4$
do fingers of God significantly degrade the discrimination between models.
In $\delta\rho / \rho > 120$ halo catalogs, the degradation is significant even
at
$F_V = 2$.  Thus lower densities increase the contribution due to
fingers of God.
Note: Values for different models are slightly offset in $F_V$ to improve
visibility.}

\begin{figure}[hp]
\plotone{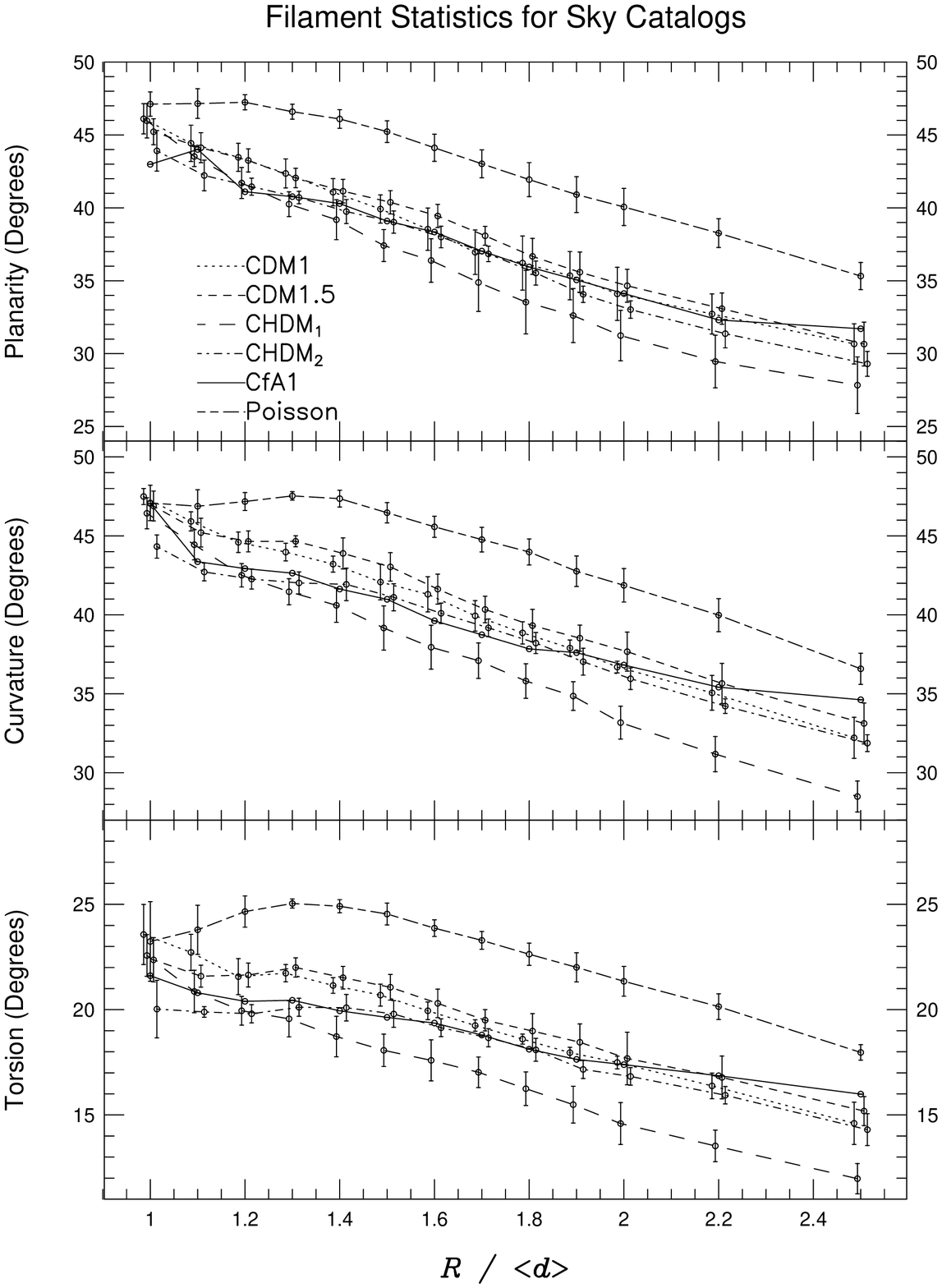}
\caption{ }
\label{fig: dgskystats}
\end{figure}

\bigskip
Figure~\ref{fig: dgskystats}.
{Filament statistics for the sky catalogs,
versus $R$ in units of ${\bar d (r)}$, the mean interparticle
spacing.  Error bars are larger than in the halo catalog statistics
due to sparseness, and Poisson is not as well discriminated from models.
CDM shows significantly less planarity, curvature, and torsion than CfA1,
while CHDM shows slightly too much.  CfA1 does not
match with any single catalog over all $R$, but does follow
CHDM$_2$ better than the other models, especially for $1.2 \leq R \leq 2.0$.
The signal-to-noise ratio between CfA1 and Poisson is highest at $R=1.3$
(note the small Poisson error bar) for all
statistics, indicating optimal sensitivity at this $R$.
Note: Values for different models are slightly offset in $R$ to improve
visibility.}

\begin{figure}[hp]
\plotone{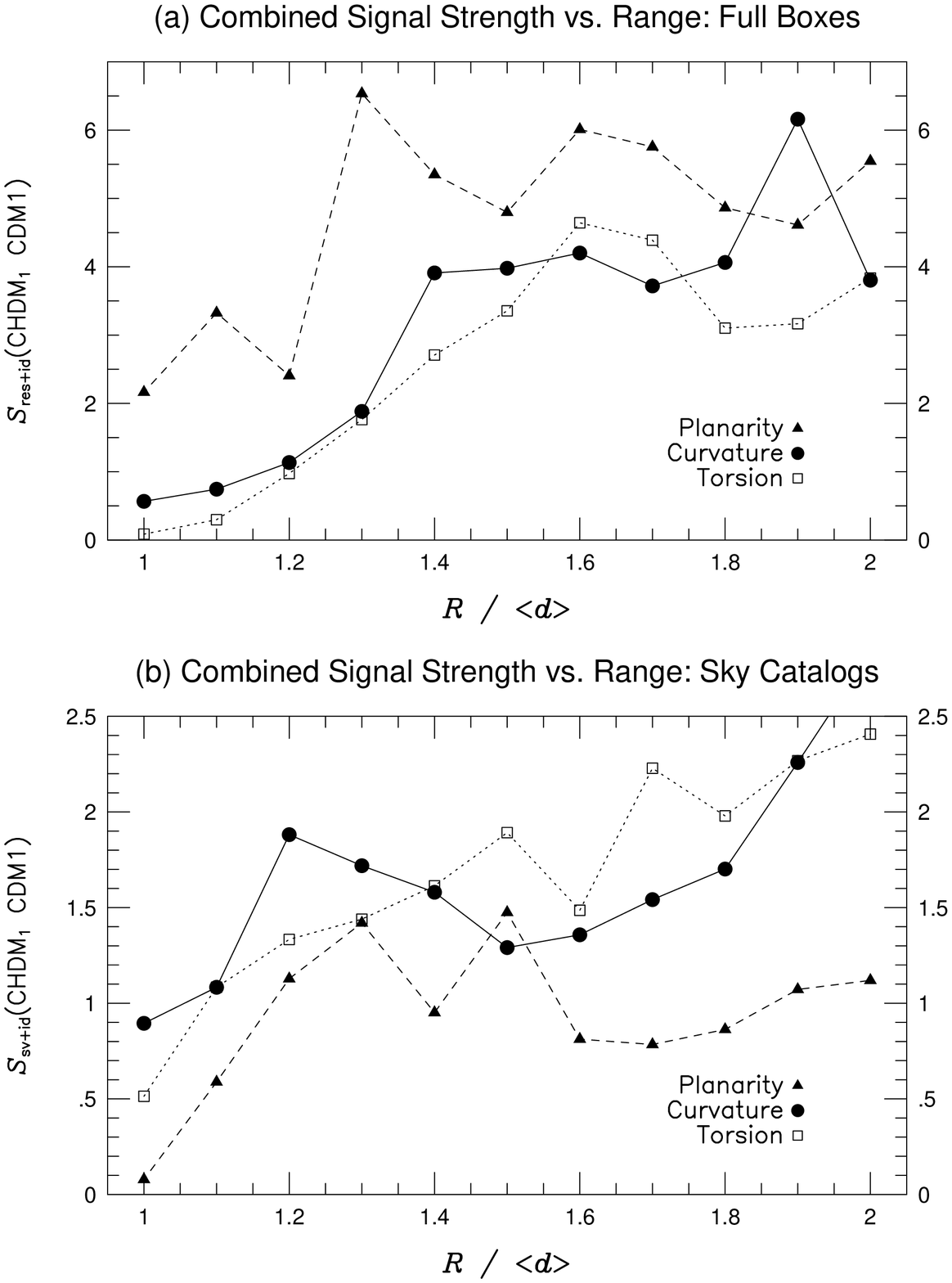}
\caption{ }
\label{fig: dgrob}
\end{figure}

\bigskip
Figure~\ref{fig: dgrob}.
{Combined signal strengths as defined in equation~\ref{eq: dgsrob};
compare to Figure~\ref{fig: dgsig} to see effect of breakup.
(a) $S^\theta_{\rm res+ID}$(CHDM$_1$,CDM1) for the halo catalogs.
Comparison with Figure~\ref{fig: dgsig}(a) shows that
breakup causes little degradation for $R\geq 1.5$.
(b) $S^\theta_{\rm sv+ID}$(CHDM$_1$,CDM1) for
the sky catalogs.  The two best statistics, $\bar\theta_C$ and $\bar\theta_P$,
drop from $\sim 2.5\sigma$ to $\sim 1.5\sigma$ under breakup in the
most sensitive range $1.2 \leq R \leq 2.0$, most likely due to added variance
from
breakup fragments' positional noise.}

\newpage

\end{document}